\documentclass[12pt]{article}
\usepackage{graphicx}
\usepackage{amssymb,amsmath,amsfonts,palatino,amsthm}
\usepackage{amssymb}
\usepackage{epstopdf}
\usepackage{slashed} 
\newcommand{\f}{\begin{equation}}
\newcommand{\ff}{\end{equation}}

\begin{document}

\title{Temporal relationalism
}

\author{Lee Smolin\thanks{lsmolin@perimeterinstitute.ca} 
\\
\\
Perimeter Institute for Theoretical Physics,\\
31 Caroline Street North, Waterloo, Ontario N2J 2Y5, Canada}
\date{\today}
\maketitle

\begin{abstract}

Because of the non-locality of quantum entanglement, realist approaches to completing quantum
mechanics have implications for our conception of space.  Quantum gravity also is expected to
predict phenomena in which the locality of classical spacetime is modified or disordered.  It is then possible that the right quantum theory of gravity will also be a completion of quantum mechanics in which the foundational puzzles in both are addressed together.  I review here the results of a program, developed with Roberto Mangabeira Unger, Marina Cortes and other collaborators,  which aims to do just that.  The results so far include energetic causal set models, time asymmetric extensions of general relativity and relational hidden variables theories, including real ensemble approaches to quantum mechanics.  These models share two assumptions: that physics is relational and that time and causality are fundamental.

Invited contribution to a collection of essays on {\it Beyond spacetime}, edited by Nick Huggett,
Keizo Matsubara and Christian Wuthrich.
 
\end{abstract}

\newpage

\tableofcontents


\section{Introduction}



I would like to introduce  a research program aimed at solving the foundational issues in quantum mechanics in a way that also addresses the problem of quantum gravity.  

There are, roughly speaking, two kinds of approaches to the measurement problem and the other issues in quantum foundations.  The first take it that the principles of quantum mechanics are largely correct and therefore aim  to make our thinking more compatible with the practice of quantum physics.  The second kind of approach begins with the hypothesis that the foundational puzzles are consequences of an incompleteness of our understanding of nature, and seeks to resolve those puzzles by means of novel hypotheses about physics.  Consequently these approaches are realist, whereas the first kind of approach are mostly anti-realist or operational.  The aim is a deeper theory, inequivalent to quantum mechanics, which completes the partial description the standard theory now reveals.  As it is a distinct theory, we may hope this completion will be testable.  The two main examples of such completions of quantum mechanics which have been studied so far, pilot wave 
theory\cite{pilot} and dynamical collapse models\cite{collapse}, do make testable predictions which differentiate them from quantum mechanics.

The program I present here is in this second class.  What distinguishes it from other realist completions of quantum theory is, first of all, the emphasis on relationalism, which the  reader may recall is a way to characterize  general relativity.  That is, we connect quantum foundations to quantum gravity by seeking a relational 
completion of quantum mechanics, i.e. an extension of the theory in terms of ``relational hidden variables."

A theory can be called relational if it satisfies a set of principles which I lay out in the next section.  These basically dictate that the beables of the theory describe relationships between dynamical actors, as opposed to defining properties in terms of a fixed, unchanging background.   But it is important to stress that there are different classes of relational theories, which differ by the choice of which of the basic elements of physics are to be regarded as fundamental,  and which emergent.  By fundamental, I mean irreducible, in the sense that it cannot be eliminated in a theory that aims to fully describe nature.    

The program I advocate here
regards time, causation, energy and
momentum as fundamental quantities.  Among the things which are emergent, and hence reducible to more fundamental elements, are space and spacetime.  Quantum mechanics itself will also be emergent from a deeper physical theory. In the next several sections I explain the motivation for these choices.  They come from reflections on issues in quantum gravity, seen in the light of quantum foundations.

The first big divide among relationalist theories has to do with whether time is regarded as fundamental
or emergent.  Many relationalists, such as Julian Barbour and Carlo Rovelli, have advocated
the view that time is emergent.  I take the opposite view, that time is fundamental. My reasons will be only touched
on here;  the full argument is presented at length in two books\cite{TR,SURT} and a number of supporting papers\cite{tn,ECS1}.  These arguments were developed in collaborations with
Roberto Mangabeira Unger and Marina Cortes.

As Rovelli has emphasized, time is a complex phenomena, and we should be precise and assert exactly  which aspects are being claimed to be fundamental.  I will be more precise in section 3 below, but let me briefly say here that  the aspect of time I assert is irreducible is its activity as the generator of novel events from present events\cite{ECS1}.  This activity generates a {\it thick 
present,} by which is meant that two events in the present can be causally related with each other.
This thick present is continually growing by the addition of novel events. At the same time other events in the thick present, having exhausted their potential to directly influence the future, slip from the present to join the always growing past. This continual construction of the future from the present, which then becomes past, makes the distinction between past, present and future objective and universal.

I am aware of the arguments that claim that an objective, observer independent distinction between the past, present and future, conflicts with special relativity.  This is addressed momentarily.

Because it is grounded on a version of relationalism that takes time to be primary, I propose
to call this program {\it temporal relationalism},  as opposed to {\it timeless relationalism}, which
supposes time to be  emergent from a timeless fundamental theory.

Once one has decided that time is going to be fundamental or emergent, one must ask the same question with regard to space.  I will explain why it is unlikely that, within a realist framework, time and space can both be fundamental.  Since I choose time to be fundamental, I must choose space to be emergent.  It is, by the way, the fact that in this construction time is fundamental, but space is not, that resolves the apparent conflict of having an objective present with special relativity.  The relativity of inertial frames, and the consequent lorentz invariance, is an emergent symmetry, which comes into effect only when space emerges, and is not hence a symmetry of the fundamental laws, which govern a domain of events with causal relations, but no space.

The third choice one has to make concerns energy and momentum: are they fundamental or emergent?  The more I reflect on the structure of our physical theories, the more I realize that momentum and energy are at the heart of
the foundations of physics, and that this distinguishes physics, in ways that are sometimes under-appreciated, from other fields of science which describe systems that change in time.  Only physics has the canonical structure indicated by $\delta \Gamma = p_a \dot{x}^a$, and the related principles of inertia and of the relativity of inertial frames.  These lead to the  most fundamental structure in quantum mechanics, the Heisenberg algebra,
$[\hat{x}, \hat{p} ] = \imath \hbar $, which cannot be expressed in a system constructed with a finite
number of qbits.  

This is one reason I find the claims to analogize nature to a computer, classical or quantum, to be inadequate: such claims neglect the fundamental roles energy and momentum each play in the structure of our physical theories, indeed, I would suggest, in nature. For reasons I describe below, I think it is interesting to develop the idea that energy and momentum  are both fundamental.

Thus, the class of theories I will develop treat time, energy and momentum as fundamental.   Space--and hence spacetime and its symmetries--are to be recovered as emergent.
Towards the end  I will describe a class of theories we constructed and studied with Marina Cortes, which realize these ideas, called {\it energetic causal set models \cite{ECS1}-\cite{ECS4}.}  

I learned about relationalism from Julian Barbour, from a conversation we had in the summer of 1980, in which he explained to me the debt that Einstein owed to Leibniz and Mach, as well as his detailed understanding of the hole problem and the role of active diffeomorphisms.  I realized immediately that I had always been an instinctive, if very naive, relationalist.  
Talking with Julian I became a 
slightly less naive relationalist.  My education has continued ever since, and I owe an enormous debt to philosophers since then. 


Temporal relationalism is a part of a larger program of {\it temporal naturalism}
which Roberto Mangabeira Unger and I presented in \cite{TR,SURT,tn}.   The adjective temporal is again meant to emphasize the centrality of the hypothesis that time is fundamental and irreducible. 

These temporal approaches are to be distinguished from timeless, or atemporal, forms of relationalism and naturalism, which hold that the most fundamental levels of description of the world are formulated without time, as time, in the sense of its flow or passage, is held not to exist fundamentally.  According to this view, time is real, but only in the sense that it is emergent, in the same sense that pressure and temperature are real.
The block universe is an aspect of timeless naturalism, so is the notion that the laws of nature are fixed and unchanging.

In the next section I introduce principles for relationalism; the primacy of time and causation is the subject of section 3.  Section 4 is devoted to the question of whether energy and momentum should be treated as fundamental or emergent.  Section  5 introduces a new notion of locality, which is entirely relational.

\section{Relationalism and its principles}


We begin with the principle behind relationalism, which is 
Leibniz's {\it Principle of sufficient reason.}
In his Monadology\cite{Monadology}, he states,

\begin{quotation}

{\it 31. Our reasonings are based on two great principles, that of contradiction, in virtue of which we judge that which involves a contradiction to be false, and that which is opposed or contradictory to the false to be true.

32. And that of sufficient reason, by virtue of which we consider that we can find no true or existent fact, no true assertion, without there being a sufficient reason why it is thus and not otherwise, although most of the time these reasons cannot be known to us.}

\end{quotation}

Leibniz also applied the principle to events\cite{events}

\begin{quotation}

{\it [T]he principle of sufficient reason, namely, that nothing happens without a reason. }

\end{quotation}

I read this to say that every time we identify some aspect of the universe which seemingly may have been different, we will discover, on further examination, a rational reason why it is so and not otherwise.  

I take this principle, not as metaphysics but, in a weak form, as methodological advice for physicists hoping to make progress with our understanding of nature.

\begin{itemize}

\item{} {\it Seek to progress by making discoveries and inventing hypotheses and theories that lesson the arbitrary elements of our theories.   } 
\end{itemize}

 I call this the principle of {\it increasingly sufficient reason} and it will be the form I employ here.

Here are some examples of steps in the history of physics which exemplified this advice.

\begin{itemize}

\item{} Eliminate references to unobservable absolute positions and motions and replace them by measurable relative positions and motions.

\item{}Seek a theory of light in which its speed is not arbitrary but is computable in terms of known constants.

\item{} Formulate the dynamics of general relativity and gauge theories directly in terms of gauge and active diffeomorphism invariant observables.  This gives us unique dynamics to leading order in the derivative expansion.

\end{itemize}

And here are some further steps which we might yet be able to accomplish.

\begin{itemize}

\item{} Seek a theory that only exists or makes sense if the number of (macroscopic) spatial dimensions is three.

\item{}Seek a completion of the standard model of particle physics that reduces the number of freely tune-able parameters.

\end{itemize}

I don't know whether or not there is an ultimate understanding in which sufficient reason is maximally satisfied and cannot be improved upon.  But each of these steps has been or would represent profound progress.  


The principle of increasingly sufficient reason has a number of consequences which are each worthy principles in their own light.  In each case they are to be read as advice, i.e. of the kinds of theories 
we ought to be aiming for, if we hope to progress fundamental physics\footnote{Notice that this advice, when followed, generates a dynamics for our theories to be challenged, and improved,
which is present in the absence of anomalies or conflicts with experiment.  This does not mean experiment plays no role for, as Feyerabend pointed out,  new theories may suggest new experiments, or give a new significance to old experiments, which can distinguish them from older theories.}.

\begin{enumerate}   
\item{}	   The principle of increased background independence.  

\item{}   The principle that properties that comprise or give rise to space, time and motion are relational.

\item{}	   The principle of causal completeness.

\item{}	   The principle of reciprocity.

\item{}	   The principle of the identity of the indiscernible.

\end{enumerate}    

Each of these requires some elaboration.

\subsubsection*{Background independence\cite{BI}}

All physical theories to date depend on structures which are fixed in time and have no justification, they are simply assumed and imposed. One example is the geometry of space in all theories prior to general relativity. In Newtonian physics, the geometry of space is simply fixed to be Euclidean three dimensional geometry. It's arbitrary, it doesn't change in time, it can't be influenced by anything. Hence it is not subject to dynamical law. 
The principle of background independence requires that the choice is made not by the theorist, but by nature, dynamically, as a part of solving the laws of physics. 

The principle states that {\it a physical theory should depend on no structures which are fixed and do not evolve dynamically in interaction with other quantities.}

Non-dynamical, fixed structures define a frozen background against which the system we are interested in evolves. I would maintain that these frozen external structures represent objects outside the system we are modeling, which influence the system, but do not themselves change. (Or whose changes are too slow to be noticed.) Hence these fixed background structures are evidence that the theory in question is incomplete.

In many cases the observables of a background dependent theory describe how some quantity changes, or some body moves, with respect to those fixed external structures.  This is the role of reference frames, they implicitly reflec a division of the universe into two parts-a dynamical system we aim to study, and a part whose dynamics we  neglect and fictionalize as fixed-for the purpose of pretending that relational quantities like relative position and relative motion, have an absolute meaning.

It follows that any theory with fixed external structures can be improved if the external elements can be unfrozen, made dynamical, and brought inside the circle of mutually interacting physical degrees of freedom. This was the strategy that led Einstein to general relativity. The geometry of space and time is frozen in Newtonian physics and it is also frozen in special relativity. In these theories the spacetime geometry provides an absolute and fixed background against which measurements are defined.  These are reflected in the role played by  reference frames. Mach pointed out the fiction involved by identifying inertial frames with the ``fixed stars."  General relativity unfreezes geometry, making it dynamical. This  freed the local inertial frames from an absolute and fixed dependence on the global mass distribution, and made their relation dynamical, subject to solutions of the equations of motion. 

This is rarely a one step process.  Typically,  the new theory retains frozen elements, then it too is not the end of our search, and will require further completion.The principle asserts we should seek to make progress by 
eliminating fixed background structures by identifying
 them as referring to external elements, and making the choices involved subject to dynamical law.

 It follows that the only complete theory of physics must be a cosmological theory, for the universe is the only system which has nothing outside of it. A theory of the whole universe must then be very different from theories of parts of the universe. It must have no fixed, frozen, timeless elements, as these refer to things outside the system described by the theory, but there is nothing outside the universe. A complete cosmological theory must be fully background independent. 

It follows that quantum mechanics cannot be a theory of the whole universe because it has fixed elements. These include the algebra of observables and the geometry of Hilbert space, including the inner product. 

This implies that there is no wave-function of the universe, because there is no observer outside the universe who could measure it. The quantum state is, and must remain, a description of part of the universe. Relational quantum mechanics develops this idea.

We then seek to complete quantum theory by eliminating background structures. We do this by exposing and then unfreezing the background and giving it dynamics. In other words, rather than quantizing gravity we seek to gravitize the quantum. We mean by that identifying and unfreezing  those aspects of quantum theory which are arbitrary and fixed, making them subject to dynamical laws.  Turning this around, we hope to understand the challenging features of quantum physics as consequences of separating the universe into two parts: the system we observe, and the rest containing the observer and their measuring instruments.
Closely related to background independence is another key idea: that the observables of physical theories should describe relationships.

\subsubsection*{Relational space and time}  

A relational observable, or property, is one that describes a relationship between two entities. In a theory without background structures, all properties that describe space, time and motion should be relational. Background independent theories speak to us about nature through relational observables.

Leibniz, Mach and Einstein taught us to distinguish absolute notions of space and time from relational notions. 


A relational theory can have some be-ables which are intrinsic to individual events and processes.  I will argue that the structure of physics requires this below.

\subsubsection*{Principle of Causal Completeness}

We follow chains of causation back in time; this principle states that
all chains of causation relate back to events and properties of our single universe.

If a theory is complete, everything that happens in the universe has a cause, which is one or more prior events. It is never is the case that the chain of causes traces back to something outside the universe. 

Consequently, our theory can only contemplate a single, causally connected universe.

\subsubsection*{Principle of Reciprocity }

This principle was introduced by Einstein in his papers on general relativity and states that if an object, A, acts on a second object, B, then B acts back on A\cite{AE-reciprocity}.

\subsubsection*{Principle of the Identity of the Indiscernible.}

This states that
any two objects that have exactly the same properties are in fact the same object.

One important implication of the identity of the indiscernible is that there are no (global) symmetries.  A symmetry of a physical theory is a group of transformation on the state space which take solutions to other solutions.  Consider an arbitrary point in the state space $a$ and 
consideer also $T \circ a$
where $T$ is a transformation.  They have exactly the same properties, hence the PII requires that we identify them.  The symmetry then acts as the identity.

An example is a translation is space.  In the full description, there is a silent background, which includes the observers and the measuring instruments.  The physical subsystem under study is being translated with respect to this silent and frozen audience.  When we identify and unfreeze the silent background and bring it into the system of dynamical equations, we eliminate the symmetry.

This does not apply to gauge transformations, which take the mathematical description of a state to a different mathematical description of the same state.

Relationalism then offers an important piece of advice to the project of the unification of the elementary particles and their interactions, which  is that the deeper a theory is the fewer global symmetries it should exhibit.  This strikingly conflicts the method that inspired the grand unified theories such  as $SU(5), SO(10)$ and $E(7)$, which aimed to unify by postulating larger and larger symmetry groups.  Of course, the program of grand unification failed experimentally, because of proton decay.

On the other hand, this is something that general relativity  gets right.  As shown 
by Kuchar\cite{karel}, general relativity restricted to spatially compact solutions has no conformal Killing fields on its configuration space, and hence no global symmetries.  From general relativity we learn that global symmetries arise only as properties of certain very special, idealized solutions, where they may be regarded as signalling the existence of a frozen, decoupled background sector.  

String theory aspires also to get this right.  Particular perturbative, background-dependent string theories have global symmetries, but these are believed to represent expansions around solutions of a more fundamental, background independent string or $\cal M$ theory, which has no global symmetries.  

The first time I heard the assertion that more fundamental theories should have fewer symmetries was in a talk by Roger 
Penrose in around 1976-7, where he used it to argue that twistor theory was right to break parity and even $CP$.
I have been thinking about it ever since.  Thus, to me one of the attractive features of loop quantum gravity, in the original formulation based on Ashtekar's original variables, is that there is evidence the quantum theory does break 
$P$ and $CP$\cite{LQG-CP}.

\subsection{Examples of relationalism in physical theories}

The importance of the methodological advice to ``make our theories more relational by eliminating background structure"
is shown in several instances where it was decisive.  The role of relationalism in the passage from Newton 
dynamics to  Einstein's general theory of relativity is paradigmatic, even iconic.  So I will begin here with 
the earlier transition between Aristotle and Newton.  I believe there is a way to tell
that story that makes it also an instance where this advice was decisive.  The importance is that the move to eliminate
background structure-in particular the notions of absolute rest and absolute velocity-resulted in the discovery of
energy and momentum and the central roles they play in the structure of physical theories ever since.

\begin{itemize}

\item{}{\it Second order equations of motion and the canonical dualities.}

One of the most important instances of the transformation of background structure into dynamical degrees of freedom occurred during the transition from Aristotelian to Newtonian physics.
For Aristotle and his followers dynamics is first order in time, i.e. a velocity is a response to a force.  More generally, if $x^a$ are the relevant dynamical quantities, which lives on a configuration space, ${\cal C}$, then the dynamics is given by a 
fixed vector field, $v^a [ x ]$ on $\cal C$.
\f
\frac{dx^a (t)}{dt} = v^a [x(t)]
\ff
This law made sense to  Aristoteleans because they believe velocity had an absolute meaning.  Hence its absence, rest, has an absolute meaning.  Newton  identified the absolute frame of reference needed to define rest as the laboratory.  He removed this background structure, by replacing absolute velocity with respect to the laboratory with its relative counterpart.  By making that subject to dynamical law he made  the decisive step to the new dynamics.  In
retrospect the key insight was the discovery of momentum, as the dynamical quantity related to relative velocity.

Most applications of mathematics to science, such as models of economics, ecology, biology etc. are still Aristotelean.  This is because they still have notions of absolute change and  stasis.  Consequently, they have no analogue of the 
principles of inertia, or of relativity.
The dynamical equations in economics, biology, ecology, general dynamical systems, etc. are thus first order in time.  Computers, and algorithms generally, such as cellular automata, have the same first order structure, in which change is given by a first order update rule
\f
X(n+1) = F[X(n)]
\ff

The fixed value of $v^a$ or $F[X(n)]$  codes the knowledge we have of the causes of change in these systems.  It is background structure. 

This goes for computers as well.  Whether classical or quantum, a computer has a well defined notion of stasis: the update rule, or the identity operation, that changes nothing.  So long as it does so, it misses a fundamental aspect of dynamics in physics,  whose deepest principle is that stasis and change are, to first order, relative.  This why I
am not convinced by claims that physics could be modelled or envisioned as a bunch of interconnected qubits.

Newton made the background structure, implicit in the fixed vector field $v^a (x)$, dynamical.

The clearest way to see this is in the Hamiltonian formalism, where the equations of motion are
\f
\dot{x}^a = \{ x^a , H \} = \frac{1}{m} g^{ab} p_b, 
\ff
But where $p_b$ rather than being fixed, has been made to evolves from arbitrary initial conditions,  subject to dynamical law.
\f
  \dot{p}_a = \{ p_a , H \} = -\frac{\partial V}{\partial x^a}
\ff

If we eliminate the $p_a$ we get the usual form of Newton's laws,  which are second order in time, so that,
\f
\frac{d^2 x^a (t)}{dt^2} = G[x(t), \frac{dx (t)}{dt} ]
\ff
so the initial velocity, $\frac{dx^a (t)}{dt} $,  rather than being a fixed background structure, is a contingent initial condition, which can be varied so that the same law applies to a multitude of different circumstances.


The canonical first order form of dynamics is elegantly defined  on a phase space, $\Gamma = (x^a, p_b )$.  This carries an imprint of the original second order form of the dynamics in the canonical pairings
\f
\{x^a, p_b \} = \delta^a_b
\ff
and in the form of the Hamiltonian
\f
H=  g^{ab} p_a p_b + V(x)
\ff
There are still elements of background structure-they are coded into the Hamiltonian in the
$g^{ab}$ and $V(x)$.  There is also the background time, which can be eliminated by
making the Hamiltonian into a constraint.  One way to do this was pioneered by Julian Barbour together with Bruno Bertotti,  Instead of the usual Lagrangian
\f
S = \int dt \sqrt{g} \left (
\frac{1}{2} g_{ab} \dot{x}^a \dot{x}^b - V(x)
\right )
\ff
we write the time reparameterization invariant lagrangian
\f
S = \int dt \sqrt{
\frac{1}{2} g_{ab} \dot{x}^a \dot{x}^b  V(x) }
\ff

I want to emphasize that the canonical form of dynamics, based on the dualism of a pair
$(x^a, p_b )$ tells us something fundamental about physics, and distinguishes motion in space from all other forms of change studied by the different sciences.

A key consequence of this structure is Noether's theorem, which relates symmetries 
in the $x^a$'s  to conservation laws involving total values of the $p_a$'s.  These symmetries
arise from invariances of  $H$ which reflect the principles of relationalism, i.e. that the forces are functions of relative coordinates.  

All of this structure is special to the laws of motion and field equations of physical systems.  

\item{}{\it Mach's principle} was a step in articulating how space, time and motion in a relational world would differ from Newtonian physics, with its dependence on absolute space and time.  It gives a criteria for a relational theory, which is that the response to an acceleration or rotation of a body, measured against ``the fixed stars" must be the same as if that body were fixed and the universe accelerated or rotated in the opposite sense.  

General relativity with spatially closed topology satisfies a form of Mach's principle.  But general relativity with asymptotically flat boundary conditions does not, because asymptotic infinity defines a fixed inertial frame of reference.

\item{}{\it Connections and gauge invariance.} One of the most basic things we know about nature is that the laws of physics are second order in time.  This means that the law of motion proscribes accelerations as a function of positions and velocity.  To define such a law we have to be able to compute the rate of change of a rate of change or, equivalently, compare two velocity vectors at slightly different times.  

To compare two vectors at different events in spacetime we need to be able to bring them together at the same point.  To do this requires a notion of parallel transport.  If we give a fixed, non-dynamical rule for comparing vectors at different events, we impose background structure, and hence violate the principle of background independence.  To satisfy that principle, we must introduce a notion of parallel transport that can be contingent and subject to dynamics.
The structure needed to do so is a connection.

The proof that a connection implements background independence is that one can rotate the tangent space at two nearby points independently, without affecting the laws of motion.  This is an example of how background independence implies an important consequence:

\item{}{\it Symmetries should all be local.}  

Indeed, it follows that a relational theory should have no  global symmetries.  This is true of general relativity
with cosmological boundary conditions\cite{karel}.  Another way to say this is that only proper subsystems of the universe can have non-vanishing values of conserved charges.  The values of 
any such charges code relations between the subsystem which is modelled and the 
rest of the universe.  The symmetries such charges 
generate translate or transform the subsystem with respect to the rest of the universe.  

\item{}{\it General relativity, with compact spatial slices} is a relational theory of space and time.
It is a story of events, their causal relations and their measure, but it is crucial to understand
that physical events do not correspond to mathematical points in a differential manifold.  Instead physical events correspond to equivalence classes of those points under active spacetime diffeomorphisms.  This means that events have no intrinsic, non-dynamical labels.  What distinguishes, and hence labels a physical event is a conjunction of degrees of freedom.  One way to say this is that physical events are labeled relationally in terms of what can be seen from there.  

Temporal relationalists disagree with timeless relationalists on one point of the interpretation of general relativity.  Timeless relationalists take the diffeomorphisms in the above discussion to include the whole group of diffeomorphisms of the four dimensional manifold.  This is in tension with some aspects of the primacy of time, in particular with the insistence on an objective notion of the present moment, and an objective distinction between past, present and future.

In classical general relativity this amounts to the claim that there is a physically preferred slicing.  Temporal relationalists then are interested in a reformulation of general relativity known as shape dynamics.  

\item{}{\it Shape dynamics} 

This is a classical gravitational theory which invokes a preferred slicing of spacetime, which arises from constant mean curvature gauge of 
general relativity\cite{SD-book}.  But the many fingered time symmetry is not broken, rather it is traded for local scale invariance  of the
Hamiltonian thee dimensional theory.  It has the same massless spin two degrees of freedom of general relativity and is equivalent to GR outside of horizons, but can differ within horizons.  

Shape dynamics is important because it shows us that all the empirical successes of special and general relativity can be made compatible with the postulation of a global time, which  would be a consequence of an objective distinction between past, (a thick) present and the future.  

\item{}{\it Entanglement} reveals that a pair of quantum systems can share properties which are not properties of either.  A completely  relational description of a quantum system would be one that constructed all its properties from such shared properties.  This is the case with the EPR state, which has no definite value of any component of the individual spins.

\item{}{\it Spin networks} are a model of a discrete or quantum geometry invented by Penrose in the early 1960s.  A Penrose spin network\cite{Roger-SN} is a trivalent graph, whose edges are labeled by half-integers (denoting representations of $SU(2)$), such that the triangle inequality is satisfied at each node.  This condition means that there is an invariant or singlet in the product of the three spin representations at each node.

Penrose was interested in the idea that the shared properties of entangled quantum systems provided a realization of Mach's principle, in the sense that the properties of  a  local system are determined by its coupling to the larger system in which it was embedded.  Mach's particular suggestion was that local initial frames are determined to be those which don't rotate or accelerate relative to the large scale distribution of matter (which he referred to as the fixed stars.).

Penrose realized a version  of Mach's principle in his spin networks. An edge in the network is chosen to be have its spin measured, along an axis which is determined by its coupling into the network as a  whole.  Through a combinatorial calculation on the graph Penrose defines an angle between a pair of spins, each on an edge. He then shows that in the limit of a large
and complex graph this reproduces the geometry of a two sphere, which is the space of directions in $R^3$.

\item{} {\it Relational hidden variable theories.}

One way to complete quantum mechanics would be by the introduction of additional degrees of freedom, whose statistical fluctuations are responsible for the fluctuations and uncertainties in quantum physics. These are called hidden variables, because it was presumed that they are not measured by the procedures which measure quantum observables.  Given the ubiquity of non-locality in quantum systems, due to entanglement, it is natural to hypothesize that the hidden variables are relational, in the sense that they describe shared properties of pairs of quantum systems. 

We then might expect to have a hidden degree of freedom for every pair of ordinary degrees of freedom.  These could be arranged as a matrix or a graph.

Several such relational hidden variables have been constructed\cite{rhv1,rhv2,rhv3}.  
These were formulated by constructing an explicit stochastic process that, in a certain large $N$
limit, the assumptions Nelson\cite{Nelson} imposes on a stochastic process to yield a solution to the
Schrodinger equation would be realized. This required a fine tuning to keep the dynamics time reversible, as discussed in \cite{rhv4}.  

\begin{enumerate}

\item{}{\it $2$ dimensional many body system\cite{rhv1}.}   The degrees of freedom are complex
numbers $Z_{ij}$ which are entries in an $N \times N$ matrix, $Z$.  The $N$ complex eigenvalues of $Z$, labeled $\lambda_i$ give the positions of $N$ particles in the
two dimensional space ${\cal C}= {\cal R}^2$.

A dynamics was imposed on the $Z_{ij}$, including an arbitrary potential energy which is a function of differences, $V[\lambda_j -\lambda_k ]$.   such that in the limit of large $N$ the eigenvalues $\lambda$ move according to Newton's laws in a potential $V$. There are
corrections that can be computed to order $\frac{1}{\sqrt{N}}$ and it is found that the probability distribution evolves according to the Schrodinger equation.

\item{} A similar relational hidden variables model was constructed for a system of $N$ particles in $D$ dimensions, where the hidden degrees of freedom are now $M$ $N\times N$ anti-Hermitian matrices\cite{rhv2}.  The eigenvalues again give the positions of the particles in $R^D$.

The  dynamics is chosen to be the standard matrix model dynamics used in dimensionally reduced models of $\cal M$ theory and Yang-Mills theory and it is again shown that the Schrodinger equation is recovered to first order in an expansion around the large $N$ limit.
                     
 \item{}A relational hidden variable whose hidden variables are a graph was constructed 
 in \cite{rhv3};
 matrix hidden variables theories were also constructed by 
 Steven Adler\cite{Adler} and Artem Starodubtsev\cite{Artem}. 

\end{enumerate}




\item{\it Relational approaches to quantum gravity}

The  search for the correct quantum theory of gravity has in my view reached an unexpected situation which is that several different approaches have achieved results which may be construed as descriptions of plausible, but distinct, regimes of quantum 
gravitational phenomenon\cite{ls-4p,missing}.  At the same time, most have also encounter significant barriers which make it seem unlikely  that it by itself is a complete theory.

Several of these approaches reflect one or more aspects of the principles of relationalism, typically by having  background independent kinematics and dynamics, in the narrow sense that their formulations do not require a fixed, classical background spacetime. The background independent approaches include causal set models\cite{cs}, 
causal dynamical triangulations\cite{cdt}, group field theory\cite{gft}
and loop quantum gravity, including both its Hamiltonian and path integral (or spin foam) formulations\cite{lqg}.

Non-background dependent formulations include perturbative string theory\cite{st}, 
asymptotic safety\cite{as}
and other perturbative approaches.

However, recent very promising  explorations of the idea that space may be emergent from entanglement (an idea which at least toughly goes back to Penrose's early work on spin networks) certainly have a relational flavour\cite{QI-holo}.

\end{itemize}

\section{The primacy of time and causation}

There are several ways of expressing the idea that time and causation are fundamental and
primary\footnote{For more discussion on these, see \cite{TR,SURT,tn,ECS1}.}.

\begin{itemize}  

\item{}The present moment is fundamental, meaning that there is no deeper level of description that lacks a present  moment, which is one of a flow of such moments.  The reason that our experience is structured as a flow of moments is because that is how the world is structured.

\item{}The distinction between the
past, present and future is objective and universal. But the present may be thick, which means it contains events that are causally related.

\item{}The process that brings further novel events from present events is real and fundamental.  This defines a thick present of events that have yet to give rise to all their future events.

\item{}The past was real and is no longer real.

\item{}The future is not real and is, to some extent, open.  There are no present facts of the matter concerning future events.

\end{itemize}

A concrete illustration of a thick present is given in the energetic causal set models we constructed
with Marina Cortes\cite{ECS1,ECS2}.  The history of the universe is a set of events with causal relations and other properties; for example each event is endowed with energy and momentum.  I can illustrate the thick present with just the causal relations.

Each event, $e$ has a causal past, ${\cal P}(e)$, which has some mathematical representation.  
There is a measure of the distinctiveness of two pasts, which we can indicate symbolically
\f
{\cal D}(e,f)= g [{\cal P}(e),{\cal P}(f)]
\ff
The causal set grows by a series of distinct moves.  In each move a new event, $e$, is created, which has two parents, lets say $a$ and $b$.   These parent events are the immediate causal past
of $e$, the full causal past is
\f
{\cal P}(f) = {\cal P}(a) \cup {\cal P}(b)
\ff
including their causal relations.

Moreover each parent can have up to two children.  At every stage in the construction, the set of events with fewer than two children makes up the thick present.

There is an algorithm which determines the next event born.  We used the rule that the
two parents are the members of the thick present which have the largest distinctiveness,
${\cal D}(e,f)$, amongst all pairs in the thick present\cite{ECS1}.  Thus, at each step, one new event is created, which has at first no children and so is part of the thick present.  One or both of the parents may leave the thick present to become part of the past, if this is their second child.  

\subsection{The nature of laws}

This view has strong implications for how we think about physical laws.

If time is to be truly fundamental and irreducible, then there cannot be absolute, unchanging deterministic laws.  One reason is that if there are such laws, then any property of the state of the world at time $t^\prime$, $P(t^\prime )$ is a function of other properties at an earlier or later time $t$,
\f
P(t^\prime ) = {\cal F}[t^\prime, t ; P_a (t)].
\ff
But, if this is so, then time has been eliminated, as any property of the system at a time other than
$t$ can be expressed in terms of properties only at $t$. 

Another way to say this is that, {\it if} there is a mathematical object, $\cal M$,
 that is a perfect mirror of the
history of the universe, in the sense that every true property of the universe, and its history, corresponds to a true theorem about $\cal M$, {\it then} each instance of causation i.e. of the generation of  a causal relation, is equivalent to an instance of 
logical implication.  But this means that causation: the generation of new events from present events, is equivalent to a timeless logical implication.  So causation, the essence of time,
is reducible to logical implication, which is timeless.  

So, when we assert that time is irreducible, we are
asserting that there exists no mathematical object, $\cal M$, which is equivalent to the history 
of the universe, in the sense specified.  Indeed we can name at least on property of the universe, which is not shared by any mathematical object, which is that here in the universe it is always some present moment, which appears uniquely, i.e. at most once, in a series of such moments.

Several of our basic principles imply that laws cannot be absolute.  The usual notion of law contradicts the principle of reciprocity because laws affect the motion of matter, but how matter moves or changes has no effect on what the laws are.  Similarly, the principle of causal completion is violated because a law that effects the motion of bodies, but cannot be affected by them, is a cause outside of the universe of causes.   

Instead, we propose that laws are only meaningful for limited ranges of space and time, and that on cosmological scales, laws evolve.  We have 
propose several mechanisms by which this might 
happen\cite{evolve,LOTC,PPr}. These are briefly mentioned below.  

It is usually presumed that the fundamental laws are symmetric under time reversal.  If indeed, time is inessential and emergent, there can be no absolute difference between the past and the future.
Since we do have such a distinction,  we are free to make the hypothesis that the fundamental laws are irreversible.  Indeed we mean this in two senses, first that there is no symmetry of the laws which reverses the direction of time; second that once an event happens it cannot be made to unhappen\cite{ECS1}.

Penrose posited that a time asymmetry of the fundamental laws could be responsible for the arrow of time\cite{Roger-asym}.  The basic idea is that a time reversible law emerges at late times from a time irreversible law, but the former is subject to a restricted set of initial conditions, which are compatible with the time asymmetry of the fundamental law.  We endorse this hypothesis and we have found some results which support it\cite{ECS1}.

\subsection{The nature of space}

What about space?  If time is fundamental, what is space?  For a relationalist, it is assumed that space manifests a network of relationships,  which define relative position and relative motion.  But are these spatial relationships fundamental or emergent from a more elementary set of relations.

In general relativity, fixing a notion of time gives us a time slicing,  $M= \Sigma \times R$.  On each slice, a
relational notion of space  is given by the spatial metric, $h_{ab}$, modulo $Diff (\Sigma )$.  
If we don't fix a time gauge, we instead view the spacetime geometry as relational, corresponding to the
spacetime metric, $g_{\mu \nu}$, modulo $Diff (M)$.

In background independent
approaches to quantum gravity, these spatial, or spacetime, relations are found to be emergent from a more 
fundamental network of relations.   In the simplest theory, which is causal set theory, both causal and metric relations
are emergent from a network of discrete causal relations.  In other background dependent approaches, such as
causal dynamical triangulations, loop  quantum gravity, spin foam models, group field theories, etc, the fundamental 
discrete structure us more complicated, but it is still the case that the spatial or spacetime geometry (modulo diffeomorphisms) is held to be emergent from a more fundamental discrete structure of dynamically evolving relationships.

Thus, the lesson from quantum gravity is that space or spacetime is emergent, not fundamental.  

It is also important to note that there are clues which indicate that both space and time cannot be both fundamental.  These include:

\begin{enumerate}

\item{} Both existing realist completions to quantum mechanics, namely pilot wave theory and dynamical collapse models,  require a preferred simultaneity, and hence are in tension with special relativity.  This is the case even in pilot wave models of relativistic quantum field theory, that reproduce the lorentz invariance in the quantum statistics.  When one puts these models out of quantum equilibrium, on sees instantaneous and non-local transmission of information which takes place in a preferred simultaneity.  

\item{}The state of the art concerning models in which classical spacetime emerges from an underlying dynamical quantum geometry is that those that succeed either have a fixed boundary, or assume a prior temporal or causal order.

\end{enumerate}

This suggests that the fundamental level of description should have a global temporal order as well as a partial causal order and that space (and hence spacetime) should emerge from the network of relations existing in the fundamental description.  

At this point we may note that the work of Sorkin and collaborators has given us a well studied  model of a fundamental spacetime which has only causal relations, this is 
causal set theory\cite{cs}\footnote{Causal sets built out of intrinsic structures was developed by \cite{Cohl,CS2,Fotini1}.}.  
As candidates for a fundamental theory, these models must be credited 
with a number of unique successes, including the only genuine prediction of a cosmological constant of the right order of magnitude, from a theory of quantum gravity\cite{Sorkin-cc}.

Another important result is that causal sets can be produced by sampling lorentzian spacetimes. However it is important to note that there is an inverse problem, which is the fact that almost no causal set arises from sampling a low dimensional spacetime.  The problem is then to choose a dynamics on causal sets which induces a low dimensional and weakly curved lorentizian spacetime to emerge. Because of the inverse problem such a dynamics must suppress the typical contributions which fail to correspond to any spacetime.

\section{Energy is fundamental}

The next (and last) choice we have to make to construct a relational theory of fundamental physics has to do with the status of energy and momentum.  We might take these to be emergent
quantities that arise as a consequence of Noether's theorem.  When the emergent spacetime has symmetries these are generated by an emergent energy and momentum.  

The other choice is to presume them fundamental.  The conservation laws of momentum and energy are then to be posited ab initio.  We conjecture that there is an inverse Noether effect whereas the conservation of energy and momentum imply the emergence of spacetime, with global translation symmetries.  We see this occur in the energetic causal set models\cite{ECS1,ECS2}, where this solves the inverse problem for causal sets.

Another reason to presume energy is fundamental is that the concept is a necessary part of Jacobson's proof of the Einstein equations from the 
thermodynamics of a more fundamental theory\cite{Ted-EES}.  

\subsection{Emergent locality}

If space and spacetime are emergent concepts, so is locality.  
To illustrate this important point,  I'd like  to present a class of theories in which locality is explicitly emergent, as a consequence of the classical equations of motion.  Moreover, since locality is a consequence of dynamics, it can be modified, to be precise, it is relativised, in an exact sense I will describe.   This is then the story called {\it relative locality.}

One way to approach this subject, which was the original way we discovered it, is to think like a phenomenologist.  Imagine that we have a quantum theory of gravity; whatever it is it depends
on {\it four} (not three) dimensional parameters, Newton's gravitational constant, $G$, $\hbar$,
$c$ and the cosmological constant $\Lambda$.  Various regimes of quantum gravity are
expressible as limits of combinations of these constants.  (These include the well known limits,
such as $G\rightarrow 0$ that gives quantum field theory on a fixed spacetime background, but there are a number of others which are not often discussed; these are discussed in \cite{missing}.)

Consider the particular limit where we hold $c$ fixed and take $\hbar$ and $G$ both to zero,
\f
\hbar \rightarrow 0, \ \ \  G \rightarrow 0
\label{RL1}
\ff
but with the ratio that defines the Planck energy fixed.
\f
E_p= \sqrt{\frac{\hbar}{Gc^3}}
\label{RL2}
\ff
This is a domain of experiment that should preserve the principle of the relativity of inertial frames
that $c$ marks, but it also has a fixed (and therefor, we expect) invariant energy scale, $E_p$.
Because this is an extension of special relativity with two invariant scales the name of this story
used to be {\it doubly} or {\it deformed} special relativity.

The same limit puts the Planck length to zero
\f
l_p = \sqrt{\frac{\hbar G}{c^3}} \rightarrow 0
\ff
so there is no quantum geometry.  We then expect that the new physics will appear first as corrections to the standard equations of relativistic physics, expressed as ratios
$\frac{E}{E_p}$, where $E$ is an energy or momentum, and will thus be naturally described
in terms of modifications of momentum space.

But before we look at those modifications let us note that there is already, with $E_p \rightarrow \infty$, a formulation of relativistic
particle dynamics that takes momentum space to be fundamental, while spacetime is an emergent
or derivative concept.

Let us start with a single free relativistic particle. It has a $4$-momentum, $p_a$, which lives
on a flat momentum space, with flat metric $\eta^{ab}$ and mass $m$.  It has an action principle, based on the extremization of
\f
S^{free} = \int ds \left (
-x^a \dot{p}_a - {\cal N}{\cal C}
\right )
\ff
 The action is invariant under reparametrizations of $s$, as a result there is an Hamiltonian constraint, which is the result of varying by the lagrange multiplier $\cal N$ is
\f
{\cal C}= \eta^{ab} p_a p_b + m^2 c^4 =0 
\ff
Notice that the only geometry that is invoked is that of momentum space.
$x^a (s)$ is a momentum  to the momentum (notice also the switch from $p_a \dot{x}^a$
to $ - x^a \dot{p}_a$, this is not a typo.) 

Varying $x^a$ yields
\f
\dot{p}_a = 0
\ff
while varying $p_a$ gets us to
\f
\dot{x}^a = - 2 {\cal N} \eta^{ab} p_b
\label{xdot}
\ff
So far this describes a free particle, which means it is at rest in momentum space.

We can add more particles trivially, by adding their actions, which are so far independent.
To begin to describe physics we introduce an interaction.  The one thing we know
about interactions is that they conserve energy and momentum.  So let us consider
an interaction where particles $A+B \rightarrow C$.  
We write $p_a^A (1)$ for the end of particle's  A world line and $p_a^C(0)$ for the beginning of particle $C$'s world line.  The conservation law says
\f
{\cal P}_a = p_a^C(0) - p_a^A (1) -p_a^B (1) = 0 
\ff
We add this conservation law to the action, using a Lagrange multiplier, $z^a$
\f
S= \Sigma_{\mbox{world lines}}  S^{free} + z^a {\cal P}_a
\ff
Now let us consider the equation that comes from varying  $p_a^C(0)$.  Unlike the other points of the worldliness, this end-point variation picks up two terms and we get
\f
x_C^a (0) = z^a
\ff
varying at the other two end-points we find they are also at $z^a$
\f
 x_A^a (1) = x_B^a (1)=x_C^a (0) = z^a
\ff
Hence the lagrange multiplier $z^a$ becomes identified with the interaction point, the event where
all three worldlines meet.  

We see that, as promised, that the interaction is local, i.e. takes place at a single event, is emergent as a consequence of equations of motion.

Now let us call that event $e$ which is embedded at point $z_e^a$.  There is another event at the end of worldliness $C$, which we will call $f$, it is embedded at $z_f^a$.  Now on $C$
the momentum is constant, so we will call it $p^C_a$.  By integrating (\ref{xdot}) we have
\f
\Delta z^a_{ef} = z_f^a - z^a_e = N \eta^{ab} p^C_b, \ \ \ \ \ N= \int_C {\cal N}
\label{z-z}
\ff
Let us consider the case of massless particles, $m_C=0$.  Then we note that the
inverse of the metric we defined on momentum space gives an induced metric on
the space of $z^a$'s.  We write
\f
\Delta z^a_{ef} \Delta z^b_{ef} \eta_{ab} = N^2 p^C_c p^C_d \eta^{ce} \eta^{df} \eta_{ef}=
N^2 p^C_c p^C_d \eta^{cd} = 0
\ff

The lesson of this story is that we start out with a theory defined on momentum space, and the classical equations of motion induce and emergent spacetime, together with an embedding of the events of our process in it, on which emerges a metric.  In the case of massless particles, what we induce is actually a conformal metric.  

\subsection{Relative locality}

Now that we are used to the idea that energy and momentum are primary and spacetime and its geometry is emergent from the classical equations of motion, we can go back to the original point which was to represent quantum gravity effects which might arise in the limit (\ref{RL1},\ref{RL2}).

We want terms in the ratio of energies to $E_p$ to appear.  How are we to do this?  We can get these from modifying the geometry of momentum space\cite{RL1,RL2}.  

A continuous geometry can depart from that of flat spacetime in two independent ways.  The metric
can be changed.  The metric comes into the constraint ${\cal C}$ where the norm of the
momentum $p_a$ must be written as $ |p|^2 = D^{2}(p)$, the geodesic distance from the point
$p_a$ of phase space to the origin.  

The other structure that can be changed is the connection or parallel transport.  This is given by a derivative
operator or connection and is generally independent of the metric.  The parallel transport
comes into the conservation law.  The momentum space is no longer a vector space and we have to think about how we define the sum of two or more momenta, which we need to state the conservation  law.  We showed that a non-linear combination operator is equivalent to a connection\cite{RL1,RL2}. 

An important point is that we want to
preserve the relativity of inertial frames, hence we want to preserve the fact that there is a symmetry group on momentum space, which has $10$ generators in the case of $3+1$
dimensions. This means we want a homogeneous geometry on momentum space.

There are three independent, invariant measures that can characterize the parallel  transport in a homogeneous geometry.  These are torsion,  non-metricity and curvature.  To give leading order
quantum gravity effects, these should be set proportional to $\frac{1}{E_p}$
and $\frac{1}{E_p^2}$.   

The effects of torsion and non-metricity have been worked out.  A typical result is the following\cite{RL2}.
Consider an observer who measures a distant event, such as the one we just discussed, 
involving a photon of energy $E$, a distance
$d$ from them.   This event we will assume is local from the point of an observer local to it.
Then the distant observer will describe it to be non-local.  For example, if the event is the
absorption of a photon by an atom, and the local observer sees the photon disappear at the spacetime point where the photon encounters the atom, the distant observer will describe the
absorption as happening when the photon was still a distance $x$ from the atom,
where in rough numbers,
\f
x= d E N = d \frac{E}{E_p}
\ff
where $N \sim  \frac{1}{E_p}$ is a component of the non-metricity tensor.   

This has observational consequences.  Consider a gamma-ray burst a distance $d$ away and
consider two photons which are emitted simultaneously, according to the local observer.  Then they
arrive at the Earth with distinct times of arrivals different by 
\f
\Delta t = \frac{d}{c}  \frac{\Delta E}{E_p}
\ff
where $\Delta E$ is the energy difference of the two photons\cite{RL2}.  

In special relativity simultaneity is relativized by locality becomes absolute.  If one observer 
sees two  events to coincide at the same time and place, that is the way all observers will see it.
As we have just seen, an analysis of the possible limits of quantum gravity phenomena, shows us that this notion of locality also gets relativized.  

The relativity of simultaneity is usually taken to mean that the notion of a spatial manifold-of
space, in other words, has no place in a world described by special relativity.  It is a fiction created by the procedure that local observers use to construct a picture of what is happening around them.
Similarly, we interpret the phenomenon of relative locality as a signal that spacetime itself is an observer dependent notion, that falls apart when one looks too closely.  Just like the surface of
simultaneity is a fiction tied to the motion of an observer, the notion of a local spacetime is also
a fiction, which depends on the position and motion of the observer, as well as on the energies of the probes she uses to observe the world distant from ourselves.

This relativization of the notion of locality was uncovered as the answer to paradoxes pointed out by \cite{sab,Unruh}
in several models with an invariant, observer independent, energy scale\cite{dsr1,dsr2,dsr3}.

\section{The dynamics of difference: replacing locality with similarity of views}

In a relational account of space, there is no intrinsic property of an object associated to its location in space.  Instead, an object's position in space is a consequence of its place in a network of relationships with other objects.  What is fundamental is that network of relationships, from which position in space is emergent and, very possibly, approximate.

The network of relationships that describes a system may be illustrated by a graph where the objects are nodes and two nodes are connected by an edge if they are involved in a relationship.  The edges may be labeled by properties of the relationship. There is a metric on the graph (one of several which exist, we will see another very different one shortly.) 
$g_{IJ}$ which counts the minimal number of steps to walk from $I$ to $J$ on the graph.

An object has a view of the system she is a part of, which represents the knowledge it may of the rest of the universe; all such knowledge is a function of the relationships which tie the object to the rest of the system.   It is very useful to describe the knowledge in terms of neighbourhoods.  The first neighbourhood of an object, $A$ consists of all objects in the network one step away on the graph, plus any edges that may join them. Similarly, the $n$'th neighbourhood of $J$, ${\cal N}^n_J$,  consists of all nodes $K$ such that
\f
g_{J,K} \leq n
\ff
together with all the links in the graph between pairs of nodes in neighbourhood.  The graphs representing neighbourhoods of $J$ have an origin, or marked point, which is the object, $J$.

One way to represent the view of an object is by the sets of neighbourhoods together with their embedding maps into each other.
\f
{\cal V}_J = \{  {\cal N}^0_J, {\cal N}^1_J, \ldots , {\cal N}^n_J , \ldots \}
\ff

I want to suggest that similarities and differences of views are fundamental, whereas distance in space is emergent and approximate.   

The usual idea of locality is that two objects will interact more often, or more strongly, the closer they are in space.  But notice that two nearby objects have similar views of the rest of the universe.  Here by your view I mean, informally, what you see when you look around, i.e. the sky from your point of view.  

We can also think of this in terms of the view of an event $V_e$.  Think of the pattern of stars seen on the sky from a particular event's perspective, i.e. the pattern of incoming radiation on the sphere which is the space of directions on your backwards light cone.  Clearly there is at least a rough relationship between the distance between two events 
\f
d(e,f) = \sqrt{|e-f|^2}
\ff
and the similarity of their views.
\f
d(e,f) \approx g(V_e , V_f )
\ff
where $g(V_e , V_f )$ is a metric on the space of views.

What if distance in spacetime is only a proxy for difference of views?  What if the locality that matters fundamentally is the distance in the space of views?  This means that two events are more likely to interact when their views are similar.

There are two ways that two events can have similar views.  One is if they are events in the history of two macroscopic bodies and are close to each other in spacetime.  This is the conventional case, which I've mentioned.  The second is if the two events arise in the histories of two atoms or molecules.  These systems have few degrees of freedom, so the space of possible views is going to be in some sense small.  But these are also systems which exist in vast numbers of copies spread through out the universe. So the view of an atom or molecule can have many neighbours in the space of views.  These neighbours define ensembles of microscopic systems, which all interact with each other in spite of being spread through the universe.  I would like to suggest that it is these ensembles that quantum states refer to.  I would also suggest that the peculiarities of quantum mechanics arises from the fact that it is a course grained description of these ensembles.

This is the basic behind the real ensemble formulation of quantum 
mechanics\cite{real1,real2,views}.

To make this suggestion precise,  we need to define the distinctiveness of
the views of two objects, $J$ and $K$, which we will call ${\cal D}_{JK}$.  There are
several ways this can be defined.  The simplest is to define
\f
{\cal D}_{JK} = \frac{1}{n_{JK}^p}
\ff
where $p$ is some fixed power and $n_{JK}$ is the smallest $n$ such that ${\cal N}^n_J $ is not isomorphic to
${\cal N}^n_K$ under all maps that preserve the origin.  

Alternatively, define a metric on views, $\mu_{JK}$,  which measures their distinctiveness modulo all isomorphisms of the views that preserve the marked points.

Given a metric on the space of views, which measures their distinctiveness, we may
define the variety of the set of relations\cite{variety1}.   
This has the general form
\f
{\cal V}= \frac{1}{N^2} \sum_{J \neq K} {\cal D}_{JK}
\ff

This is, in general, an interaction amongst three systems, because for each pair $J$ and $K$ the distinctiveness ${\cal D}_{JK}$ involves a comparison of the views $J$ and $K$ each have to third bodies.  Very remarkably, this turns out to yield exactly Bohm's quantum potential\cite{real2}.

This is the core of how Schrodinger quantum mechanics emerges
from a dynamics that involves comparisons amongst similar systems.  I leave the
details to the papers\cite{real1,real2,views}.


\subsection{Causal sets and causal neighbourhoods}

Given the emphasis of the primacy of time and causation,  we will be interested in
models of fundamental physics based on causal structures.  A set of
events, $\{ A,B,C, \ldots , \in {\cal C}$ together with a  causal relation, $>$, is a causal set
if for any two events, $A$ and $B$, only one of the following is true:
\f
A >B, \ \ \ \ B >A, \ \ \ \ \mbox{or, A and B are causally unrelated}.
\ff
We also require that the set be {\it locally finite}, i.e. for all $A$ and $B$, the set of $C$ such that $B> C> A$  is finite.

We require that causal relations are transitive, so that
\f
A < B, \ \mbox{and} \ B < C \ \mbox{implies} \ A< C
\ff
We will also want to denote the primary causal relations, A <-- B, which are irreducible and generate the rest.

The view of an event is its causal past:
\f
B \in {\cal P} (A)   \forall B \ \ \  \mbox{such that} \ \ \   A > B
\ff
We can define the first, second and $n$'th causal neighbourhoods as the subsets of
${\cal P}(A)$ which are $N$ or fewer causal steps to the past of $A$.   These are denoted
${\cal P}_n (A)$

We can define the causal versions of distinctiveness.
\f
{\cal D}_{AB} = \frac{1}{n_{AB}^p}
\ff
where $p$ is some fixed power and $n_{AB}$ is the smallest $n$ such that ${\cal P}^n_A $ is not isomorphic to ${\cal P}^n_B$. 

We then define the causal version of variety
\f
{\cal V}= \frac{1}{N^2} \sum_{A \neq B} {\cal D}_{AB}
\ff

We can then consider theories based on an energetic causal set, in which the variety defined in this way acts as a potential energy.  The dynamics of such a theory aims to maximize the variety of the system.  This acts preferentially on pairs with small distinctiveness, and changes them so as to increase the overall variety.  In other words, the more similar two events are, the more likely they are to interact.  

Under appropriate conditions, this leads also to a derivation of Schrodinger quantum mechanics, from a theory whose dynamics involves extremizing the variety.  I leave the details to the original paper\cite{views}.

\section{Temporal relationalism}

As I said in the introduction, the program of temporal relationalism 
 aims to complete quantum mechanics in a way that would simultaneously construct a quantum theory of gravity and spacetime. It can aspire to do both because it is based on a series of closely interconnected hypotheses about the nature of space and time.  We discussed each of these in the preceding pages; it is now time to put them together and summarize the resulting picture.

\begin{enumerate}

\item{}Time, in the sense of causation is fundamental, by which is meant that it is irreducible\cite{SURT,TR,tn}.  The activity of time is the unceasing and irreversible generation of novel events from present events; this generates a continually growing network of causal relations\cite{ECS1}.

\item{}Energy and momentum are also fundamental and irreducible\cite{RL1,RL2,ECS1,ECS2}.  
Events are endowed with energy and momentum and are  transmitted 
by their causal relations\cite{ECS1,ECS2}. 

\item{} Space is not present fundamentally, but is emergent from the underlying, dynamically evolving network of causal relations\cite{ECS1,ECS2}.  Locality is emergent, as therefor is non-locality.  

\item{}Lorentzian spacetime geometry is an emergent, macroscopic coarse graining of the fundamental causal structure, and the Einstein equations are consequences of the
statistical thermodynamics which traces the flow of energy through the fundamental network of causal relations\cite{Ted-EES,ls-sf,ls-4p}.

\item{}Locality is disordered, in the sense that the there are mismatches between the emergent metric causal structure  and the fundamental causal structure\cite{disordered}.  These may be responsible for quantum non-locality\cite{rhv3}.

\item{}Locality in space is a consequence of locality in a space of views, where a view of an event is defined to be the information it receives from its causal past\cite{views}.

\item{}The most fundamental law is the one that 
generates new events and is irreversible\cite{ECS1}.  Other laws represent regularities, mostly emergent, and are local and evolving\cite{SURT,TR}. 

\item{} Local, reversible laws emerge from global irreversible laws\cite{ECS4}.  This explains the existence of arrows of time.

\end{enumerate}

The main results of this research program to date include the following.

\begin{itemize}

\item{}{\it Relative locality.} By studying a limit of quantum gravity defined by (\ref{RL1},\ref{RL2}), we find an extension of special relativity where causation, energy and momentum are fundamental, and space and spacetime are
emergent\cite{RL1,RL2}. Locality is relativized, so that whether two events are observed to coincide depends on their
energies and distance from the observer.

\item{}{\it Energetic causal set models\cite{ECS1}-\cite{ECS4}.}  These were constructed with Marina Cortes and describe a world in which time, causation, energy and momentum are fundamental and all take place in a spaceless world.  Some of the results include:

\begin{itemize}

\item{}The emergence of a lorentzian spacetime together with an embedding of the causal set into it\cite{ECS1,ECS2}.

\item{}The evolution passes through an initial phase, which is chaotic and irreversible, which gives way to an ordered phase where there emerge pseudo-particle-like excitations whose dynamics is quasi-reversible\cite{ECS1}.  We now understand this to be very similar to the behaviour of deterministic finite state dynamical systems, in which the initial phase is dominated by a basin of attraction which leads to a phase dominated by limit cycles\cite{ECS4}.  This mechanism can explain the emergence of time reversible effective laws from a fundamental theory which is irreversible.

\item{}In the limit of many events there emerges the dynamics of relativistic particles moving and interacting in the emergent lorenzian spacetime\cite{ECS1,ECS2}.

\item{}A class of spin foam models invented by Wolfgang Wieland\cite{WW} can be understood as
energetic causal set models\cite{ECS3}.

\item{}There appears to be a disordering of causality in the sense that the causal ordering in the fundamental law generating the causal set is not entirely consistent with the causal structure of the emergent spacetime in which the causal set is embedded\cite{inprep}.

\end{itemize}
 
 \item{}{\it Time irreversible extensions of general relativity\cite{TA1,TA2}.} If general relativity emerges from a time irreversible theory early in the history of the universe, there ought to be an extension of general relativity which is not symmetric under time reversal.  This could serve as an effective theory to describe the transition from the time asymmetric to the time symmetric dominated regime.
 We constructed two classes of such theories\cite{TA1,TA2} and studied their implications for the physics of the early universe\cite{TA3}.
 
 \item{}{\it The real ensemble formulation of quantum mechanics\cite{real1,real2,views}.}  This is a completion of quantum mechanics in which the wave-function of a microscopic system is derived from an ensemble consisting of all the similar systems spread through out the universe.  We posit a new and highly non-local interaction amongst the members of such ensembles, which extremizes the variety of the system, where by variety we mean a measure of the diversity of the views of the different subsystems\cite{variety1}.  We find that the variety is closely related to the Bohmian 
 quantum potential\cite{real2,views}.
 
 \item{}A class of relational hidden variables theories, in which the hidden variables are shared properties, associated to pairs of particles, was constructed and studied\cite{rhv1}-\cite{rhv4}.
 
 \item{}{\it Evolving laws.} One of the big questions of fundamental physics is how the universe chooses its laws from a reservoir of many possible laws.  A natural solution is that the laws evolve in time\cite{evolve,LOTC}.  This gives rise to the meta-law dilemma\cite{SURT} which starts by asking whether there is a law which governs the evolution of laws and, if so, how that was chosen.   We proposed so far three approaches to this question.
 
 \begin{enumerate}
 
 \item{}{\it Cosmological natural selection\cite{evolve,LOTC,CNS-review}.} This is a proposal I made in the late 1980's of a cosmological scenario in which laws and their parameters evolve on a landscape of possible laws (analogous to the fitness landscape of population biology.
 
 \item{}{\it The Principle of Precedence\cite{PPr}} is a proposal for laws to the dynamical laws of quantum mechanics to evolve.
 
 \item{}{\it Universality of metalaws}  A set of model metalaws are studied and there is shown to be a notion of universality of metalaws, such that one can't distinguish them from each 
other\cite{unify-model,unify-universal}.

 \end{enumerate}
 
 \item{}{\it  A causal theory of views\cite{views}}  arises by the combination of the real ensemble formulation of quantum mechanics with the energetic causal sets.  The beables are the causal views of each event.  
 
 \item{}{\it The emergence of the Einstein equations from the thermodynamics of an energetic 
 causal set.   }   Conditions are given to demonstrate the emergence of general relativity from the thermodynamic limit of an energetic causal set theory\cite{ls-sf,ls-4p} to which certain additional
 conditions have been imposed.   These
 conditions are expressed in \cite{ls-4p} as four principles, the first of which is equivalent to the statement that a quantum spacetime is described by an energetic causal set 
model.

I refer the reader to the original paper \cite{ls-4p} for a full discussion and motivation of
these principles.  The key point for our discussion in this paper is that I am able to
show that these principles suffice to recover the Einstein equations  from a certain kind of energetic
causal set model\cite{ls-4p}.  To do this, I follow a strategy pioneered by 
Jacobson\cite{Ted-EES,Ted2015} and Padmanabhan\cite{Paddy5}.  

\end{itemize}

\section{Conclusions}

What is beyond spacetime?  

I have described the main ideas, and main results to date, of a research program which aims to find out.  We postulate that time, causation, energy and momentum are fundamental and irreducible, but  space and spacetime  are emergent.  Quantum physics is also emergent from a more fundamental dynamics, based on the principle of maximal variety\cite{variety1}.  From the resulting viewpoint,  locality is discovered to be a proxy for similarity of views by which a subsystem of the universe sees their position in the network of relations and interactions that defines the world.



\section*{Acknowledgements}

I first of all want to thank my collaborators with whom the program described here was formulated: Roberto Mangabeira Unger
and Marina Cortes.  I am grateful also to Giovanni Amelino Camelia, Stephon Alexander, Henrique Gomes, 
Jerzy Kowalski Glickman, Andrew Liddle, Joao Magueijo,  and Yigit Yargic for collaborations on related work.

A life time of conversations and/or  collaborations with Abhay Ashtekar, Julian Barbour, Laurent Freidel, Stuart Kauffman, Renate Loll, Fotini Markopoulou, Carlo Rovelli, Rafael Sorkin and Antony Valentini have been essential to my intellectual life.

Dialogues with philosophers have been essential to my work on these issues, among many I am
especially grateful to David Albert, Harvey Brown, Jim Brown, Jennan Ismael, Simon Saunders and Steve Weinstein.  While I was still a student, my life was changed by encounters with two great philosophers of science, Abner Shimony and Paul Feyerabend.  I also wish to thank Nick Huggett,
Keizo Matsubara and Christian Wuthrich for the invitation to contribute to this collection.  

This research was supported in part by Perimeter Institute for Theoretical Physics. Research at Perimeter Institute is supported by the Government of Canada through Industry Canada and by the Province of Ontario through the Ministry of Research and Innovation. This research was also partly supported by grants from NSERC and FQXi.  I am especially thankful to the John Templeton Foundation for their generous support of this project.

\end{document}